\documentclass[aps,pre,twocolumn,groupedaffiliation,showpacs]{revtex4-1}
\usepackage{graphicx}
\usepackage{amsmath}
\def\be{\begin{equation}}
\def\ee{\end{equation}}

\def\bc{\begin{center}} 
\def\ec{\end{center}}
\def\bea{\begin{eqnarray}}
\def\eea{\end{eqnarray}}

\usepackage{amssymb}
\usepackage{amsthm}
\theoremstyle{definition} 
\theoremstyle{plain} 
\theoremstyle{remark} 

\begin{document}
\title{Overelaxed hit-and-run Monte Carlo for the uniform sampling of convex bodies with applications in metabolic network analysis.}
\author{G.De Concini$^1$, D.De Martino$^2$}
\affiliation{$^1$ Stanford University, Department of Electrical Engineering, 350 Serra Mall, Stanford, CA}
\affiliation{$^2$ Center for life nanoscience CLNS-IIT, P.le A.Moro 2, 00815, Rome, Italy}

\begin{abstract}
The uniform sampling of convex regions in high dimension is an important computational issue, from both theoretical and applied point of view. 
The hit-and-run montecarlo algorithms are the most efficient methods known to perform it and one of their bottlenecks relies in the difficulty of escaping from tight corners in high dimension.
Inspired by optimized montecarlo methods used in statistical mechanics we define a new  algorithm by overelaxing the hit-and-run dynamics. 
We made numerical simulations on high dimensional simplices in order to test its performances, pointing out its improved ability to escape from angles and  
finally apply it to an inference problem in the steady state dynamics of metabolic networks. 
\end{abstract}

\maketitle
\section{Introduction}
The use of concepts and techniques from statical mechanics to solve hard computational problems has been very fruitful, 
in particular the application of montecarlo methods to integrate in high dimensions has been defined a revolution\cite{1} with
examples that range from the resolution of  combinatorial optimization problems\cite{2} to the simulation of complex physical systems\cite{3}.
One simple yet non-trivial computational issue concerns the uniform sampling of convex regions in high dimension. 
This  is for instance the main step for the volume computation of convex bodies\cite{4}, a problem that can be mapped in general onto the  calculation of a permanent\cite{5,6} and thus it is \#P-hard.
Even if deterministically non-polynomial, this computation can be carried out with a stochastic algorithm in polynomial time provided a uniform sampling of the body in interest,
and this is one of the cases in which the dramatic improvement accomplished by a stochastic method can be rigorously demonstrated\cite{7,8}. 
Apart from that, an uniform sampling  of convex regions is generally required  in solving inference problems from uniform priors given by linear constraints: 
examples include recostrutions of mass distributions from gravitational lensing in astrophysics\cite{9} and the statistical analysis of stationary fluxes in metabolic networks\cite{10}.  
In  principle the uniform sampling can be done by simple interpolation if all the vertices of a polyhedron are known, 
but since their number can increase exponentially with the dimension this makes their enumeration often infeasible.
Approximate algorithms based on linear programming that find a subset of rapresentative vertices are sometimes used\cite{11,12}, 
but they do not guarantee the uniformity of the sampling\cite{9}. On the other hand, several classes of random walks have been developed in order to achieve effective  uniform sampling, 
such as the grid-walk and the ball-walk algorithms\cite{4}. The most efficient montecarlo method for sampling so far is the Hit-And-Run algorithm\cite{13,13b}. 
The Hit-And-Run algorithm has been extensively studied, focusing in particular on the convergence time 
to the stationary uniform distribution, that is polynomial in the dimension space with a bound that is the best obtained so far for convex body sampling\cite{14}.
Two bottlenecks of this algorithm regard the heterogeneity of the scales and the difficulty to escape from narrow angles in high dimension. 
The first has been the focus of many research efforts\cite{Lovazs}, whereas 
the second problem instead is less investigated since it is much improved with respect to simpler random walks\cite{16}.
However  this issue  can still affects the convergence properties and the focus of our work will be thus on this problem.   
On the other hand, one of the issues arising in the simulation of large scale disordered systems in condensed matter and statistical mechanics 
is the entrapment in metastable states, a problem that inspired the formulation of optimized montecarlo schemes\cite{17} like for instance overelaxing techniques.
These methods, originally proposed in the context of  quantum field theories\cite{18}, they were extended in the field statistical mechanics of spin systems\cite{19}, 
and recently applied to general problems of inference\cite{20}. 
The basic idea of these methods is to select, during the dynamics, a change for the current point in a region of phase space which is as far as possible from the old value.
Inspired by this strategy in this paper we introduce a new Hit-And-Run  algorithm by  overelaxing the dynamics
and thus allowing the sample points to get out faster from corners. 
In the first section, after a brief description of the original method, we will outline in detail the structure of our algorithm; 
in the second we will discuss its performance showing numerical results of extensive simulations on high dimensional simplices, pointing out  
the ability of the walk to avoid the entrapment in tight angles. Then we will show an application  in the field of constraint-based modeling of cell metabolism, 
and finally, we will draw out some conclusions summarizing the main results and pointing out several interesting future extensions for our work.

\section{The overelaxed Hit-And-Run algorithm} 
\begin{figure}[h!]\label{fig0}
\begin{center}
\includegraphics*[width=.4\textwidth,angle=270]{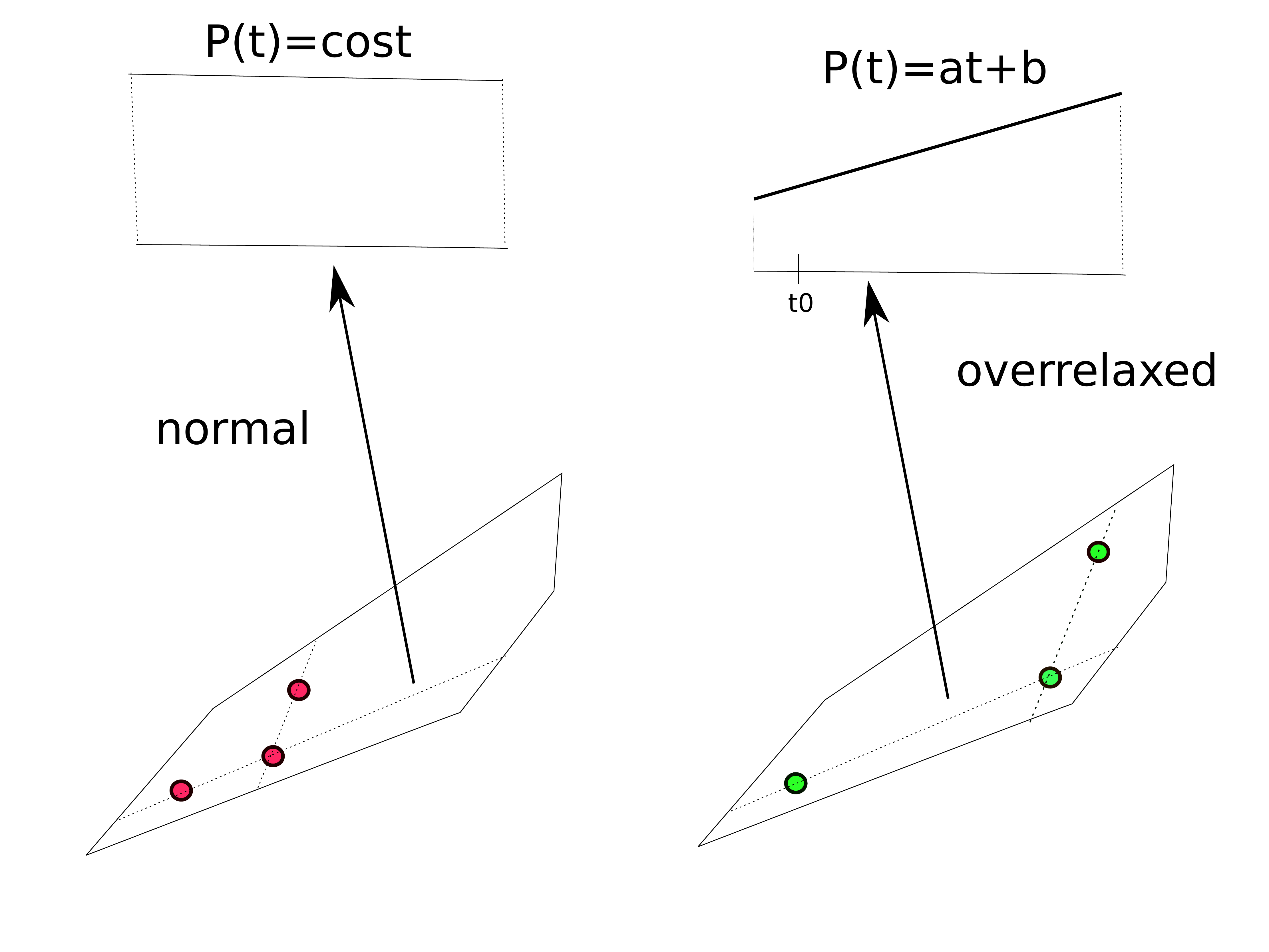}
\caption{Scheme of the difference between normal (left) and overelaxed (right) hit-and-run dynamics.}
\end{center}
\end{figure}

Given an $D$-dimensional convex set $K$, from which one wants to sample from, and a point $x_t \in K$, the Hit-And-Run algorithm is defined as follows:
\begin{enumerate}
  \item choose a uniformly distributed direction $\mathbf{\theta}_t$;
  \item choose uniformly $\lambda_t \in \Lambda_t = \{\lambda \in \mathbb{R} : \mathbf{x}_t+\mathbf{\lambda}\mathbf{\theta}_t \in S\}$
  \item move to the point $\mathbf{x}_{t+1} = \mathbf{x}_{t} + \lambda \mathbf{\theta}_t$, increment $t$ by one and start again from (i).
\end{enumerate}
The first step can be carried out with the Marsaglia method, i.e. by generating $D$ independent gaussian random variables with unit variance and then normalizing them.
In the overelaxed version of the algorithm, after drawing the segment trough the current point, we don't extract uniformly another point on this segment but
rather we sample it in an anti-correlated fashion with the current point, paying attention to mantain the detailed balance with respect to the uniform distribution.
If the subsesequent points of the markov chain are anticorrelated in such a way that their distances increase with respect to the normal algorithm they should explore the feasible space in reduced times. 
The simplest choice for the marginal conidtional distribution on the segment that we can think of is linear. 
Let's put the origin on one extrem of the segment, the other being $L$ (lenght of the segment), if the coordinate of the current point along the segment is $t_0$, a correctly normalized conditional 
marginal distribution that verifies detailed balance is (see the appendix for a straightforward derivation):
\begin{equation}
P(t|t_0) = \beta (t-L/2)(t_0-L/2) +1/L.
\end{equation}
This expression is symmetric with respect $t$ and $t_0$ and thus implement detailed balance with respect to the uniform distribution. 
Here $\beta$ rules the correlation between $t$ and $t_0$: a choice that maximizes the anticorrelation is $\beta = -4/L^3$, we have
\begin{equation}
P(t|t_0) = \frac{1}{L}(1-(2\frac{t}{L}-1)(2\frac{t_0}{L})).
\end{equation}
This can be integrated and inverted and it gives
\begin{eqnarray}\label{calct} 
t_1 = L \frac{1+A - \sqrt{(1+A)^2-4AR}}{2A} \\
A = 2\frac{t_0}{L}-1 
\end{eqnarray}
where $R$ is a random number uniformly distributed in $[0,1]$. In fig \ref{fig0} we illustrate schematically the algorithm.

The algorithm has no free parameters and its computational cost with respect to the normal algorithm requires only the evaluation of expression \ref{calct}.
Lovasz showed in\cite{14} that the simple hit and run dynamics mixes in $O(D^2 \frac{R^2}{r^2} )$ steps, 
when applied to a convex body whose circumscribed and inscribed sphere have respectively radii $R$ and $r$. 
An high value of the  factor $R/r$ stands for somehow ill-conditioned costraints that defines bodies with very heterogeneous scales 
and can be reduced in general to $O(D^{3/2})$ by appropriate preprocessing\cite{Lovazs},
the $O(D^2)$ factor is instead partially related to the difficulty in escaping from tight angles, as we will show in the next paragraph.

\section{Performance tests on simpleces}
A natural class of homogeneous convex bodies over which explore the properties of our method are the simplices $\sum_i x_i \leq	1, \quad x_i\geq 0$.
We thus measure the center of mass monitoring its dependence from the elapsed machine time on bodies of exponentially increasing dimension. 
In fig.2 we show the error as a fuction of the machine time on the first coordinate of the center of mass for simpleces of dimensions $D=4,8,16,32$ respectively (averages are taken over $2000$ samples).
The overelaxed algorithm shows a lower error with a decrease that is approximatively constant  $\delta = \frac{\sigma_{normal}-\sigma_{over}}{\sigma_{normal}}\simeq 10 \%$, 
that leads to an improvement of the samplig times of the order $1-(1-\delta)^2 \simeq 20\%$.   
\begin{figure}[ht!]\label{fig1}
\begin{center}
\includegraphics*[width=.23\textwidth,angle=0]{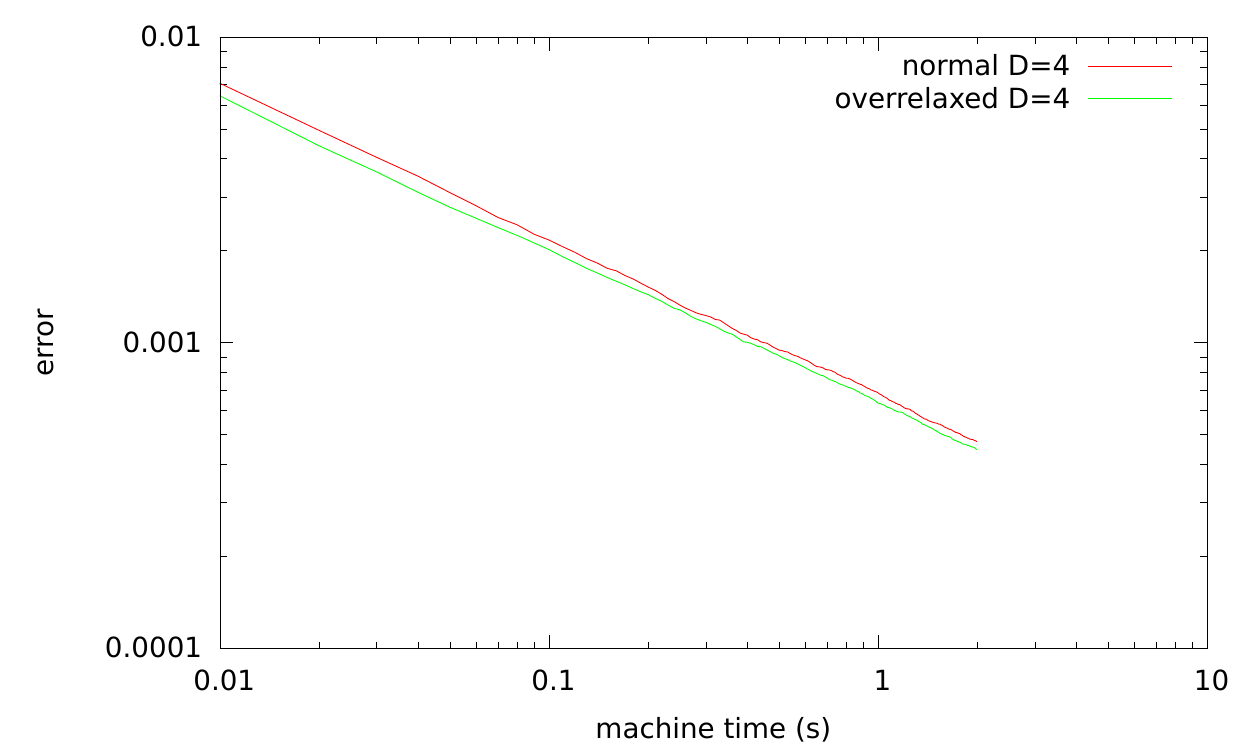}
\includegraphics*[width=.23\textwidth,angle=0]{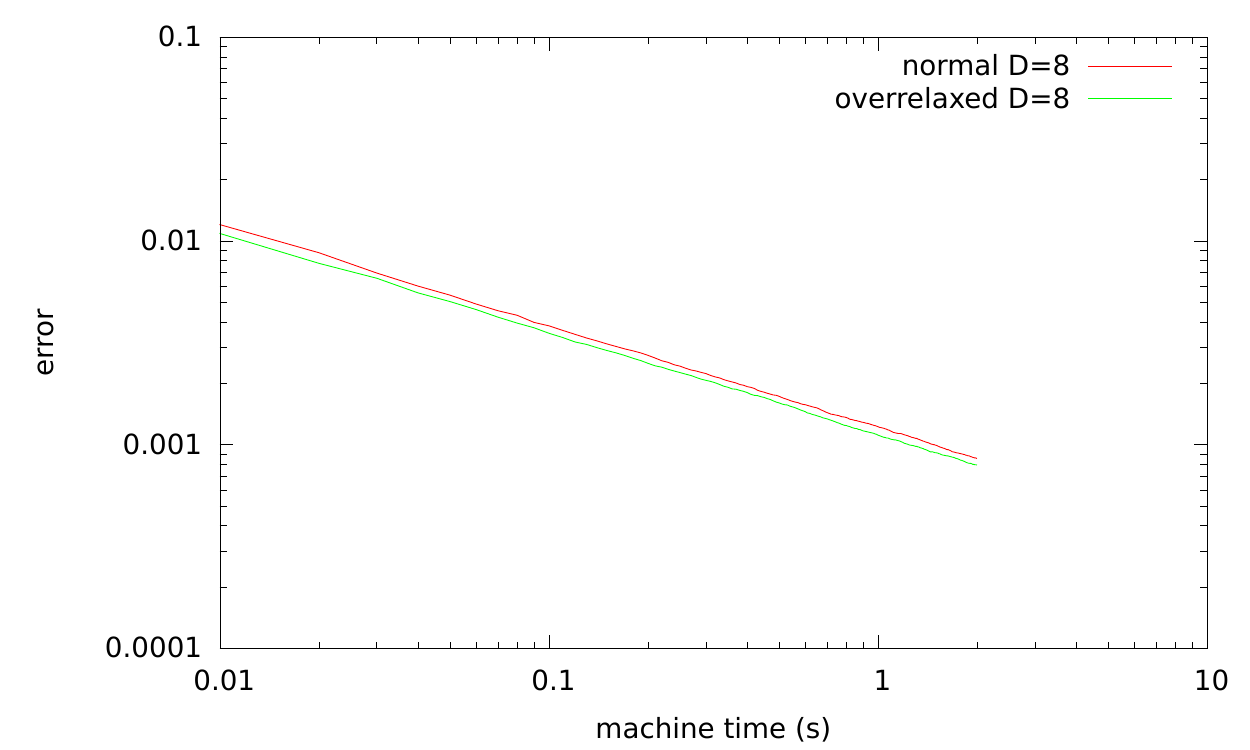}
\includegraphics*[width=.23\textwidth,angle=0]{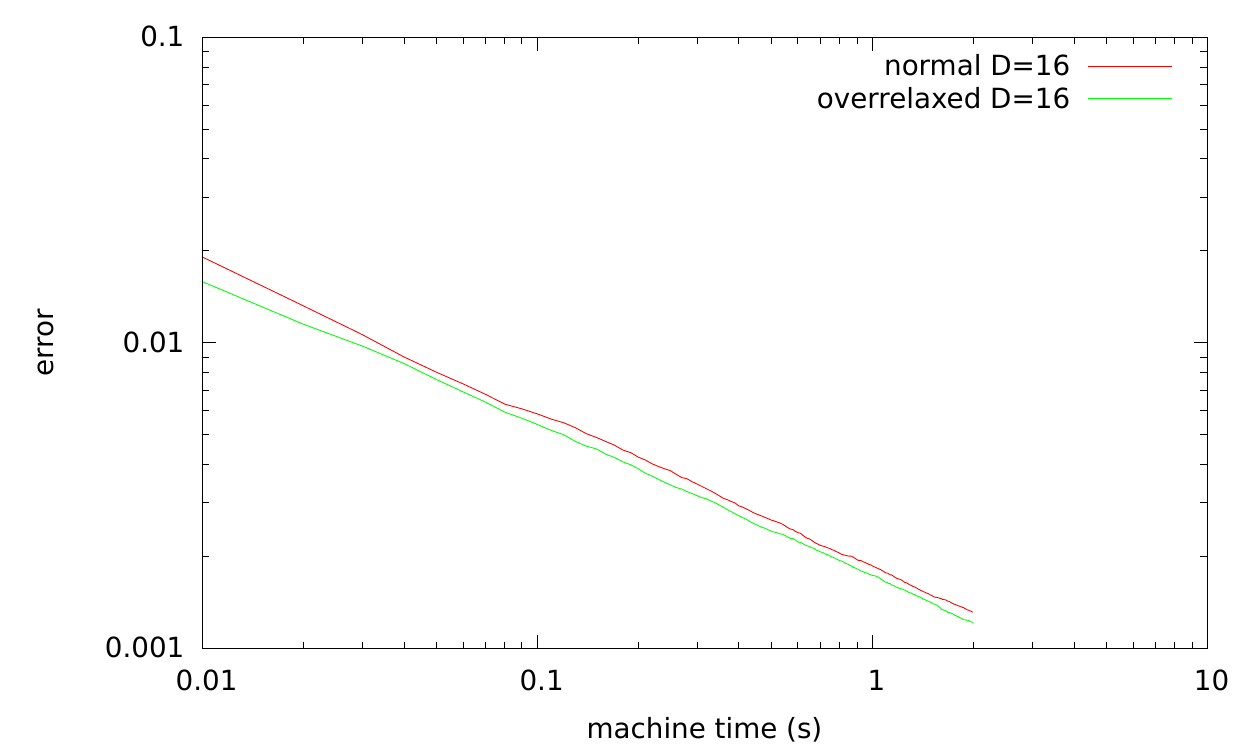}
\includegraphics*[width=.23\textwidth,angle=0]{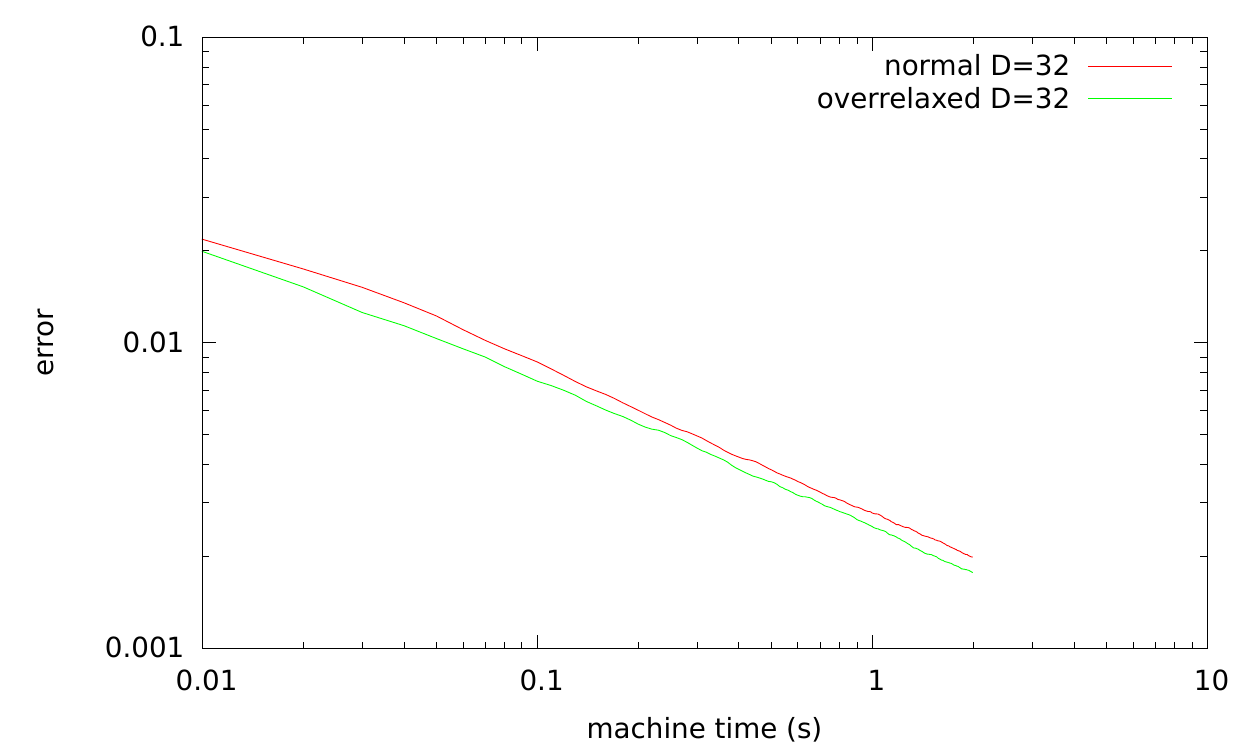}
\caption{Error as a fuction of the machine time (on an Intel dual core at 3.06 GHz) of the first coordinate of the center of mass for simpleces of dimensions $D=4,8,16,32$ for normal and overelaxed hit-and-run dynamics($2000$ samples, log-log scale).The error of the overelaxed dynamics is approximatively  $10\%$ lower than the normal, leading to an improvement of the samplig times of the order of $20\%$.}
\end{center}
\end{figure}
These performances can be partially ascribed to the improved ability to escape from angles. 
We have thus measured the escaping times from angles in simpleces in both algorithms: starting from a point in the corner with coordinates $x_i = 1-N\epsilon$, $x_{j\neq i}=\epsilon$ ( we consider $\epsilon = 10^{-6}$)
we repeteadly measure the time it takes in order that the distance from this corner becomes lower than the distance from the center of mass. 
In fig. 3  (left) we show the distribution of such escaping times (in number of montecarlo steps), that is peaked to a lower value  for the overelaxed algorithm.
In fig. 3 (right) it is possible to appraise the dependence of the average escaping time as a function of the simpleces' dimension, 
that gives the quadratic scaling typical of such hit-and-run algorithms, with a lower factor constant for the overelaxed version.
\begin{figure}[h!]\label{fig2}
\includegraphics*[width=.23\textwidth,angle=0]{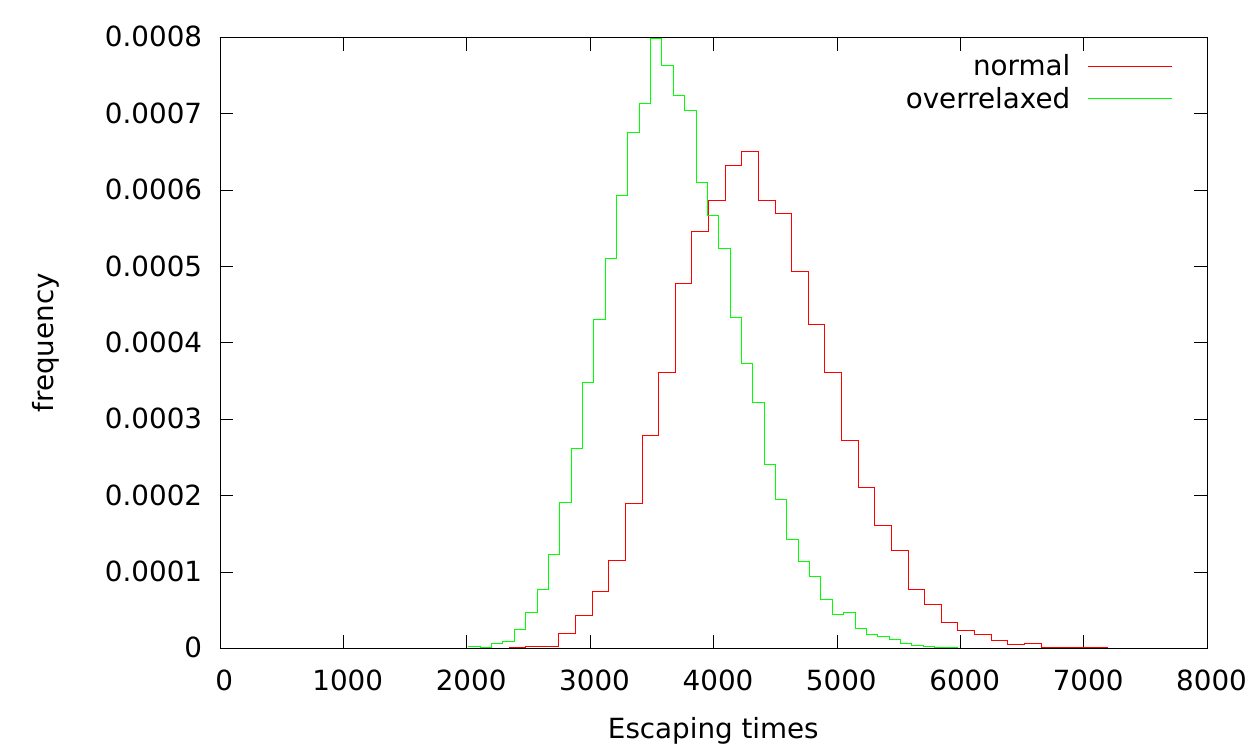}
\hspace{\fill}
\includegraphics*[width=.23\textwidth,angle=0]{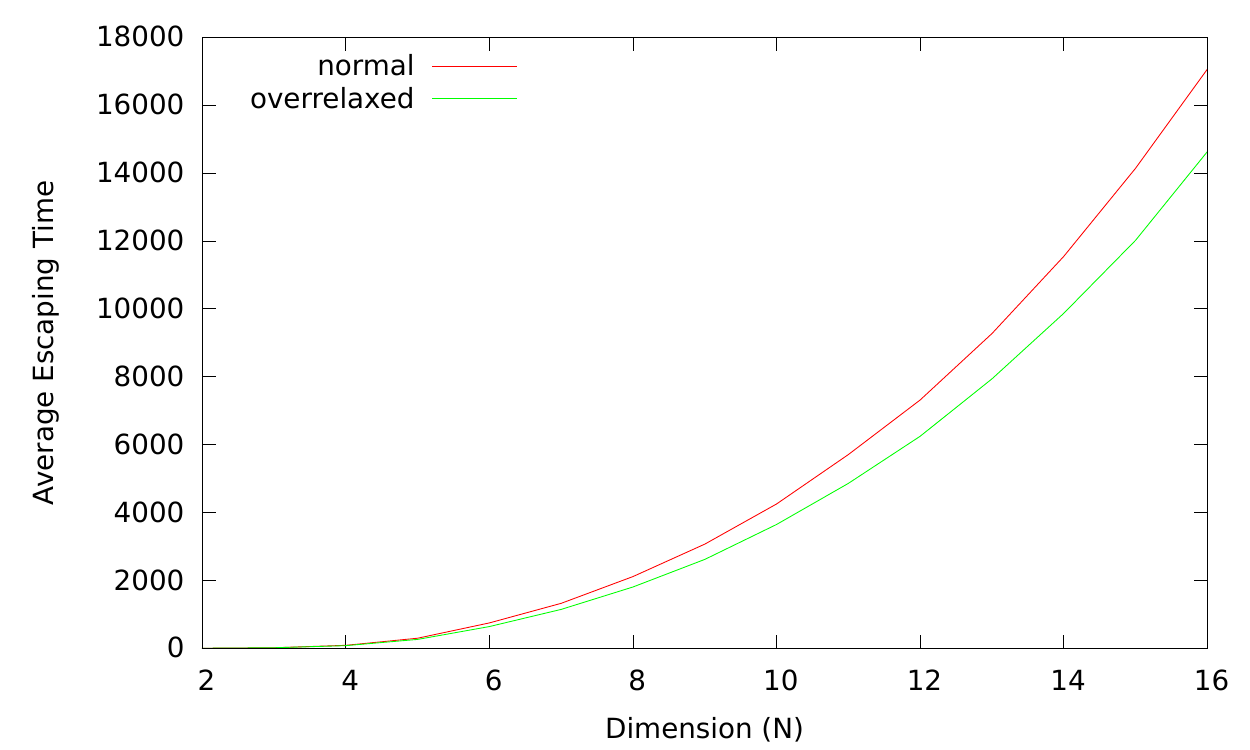}
\caption{Left: distribution of the escaping times from corners (in number of montecarlo steps) for the normal and overelaxed dynamics ($D=10$, $2\cdot 10^4$ samples). 
Right: average escaping time as a function of the simpleces dimension.
The average escaping time of the overelaxed dynamics is approximatively $20\%$ lower than the normal one.}
\end{figure}
\section{Application to inference in metabolic networks analysis}
As a matter of illustration of the application of our algorithm in the field of the inference from uniform priors defined by linear constraints 
we performed a statistical analysis of the stationary fluxes of the catabolic core of the genome-scale reconstruction of a metabolic network of the bacterium E.Coli\cite{21}. 
Cell metabolism relies on a network of enzymes that is able to degrade nutrients in order to fullfil all the cell free energy requirements. 
The simplest model of metabolism is a chemical reaction network in the steady state and it is the only one computationally 
feasible if kinetic details are unknown as it is the case for large scale instances. 
A metabolic network can be represented by a matrix $\bf{S} = (S_i^\mu)$ encoding the stochiometric coefficients of the compound $\mu$ into reaction $i$
and the stationary flux vector ${\bf v}$ satisfies the linear mass-balance equations  with prescribed bounds
\begin{eqnarray}\label{mb}
\mathbf{S \cdot v}=0, \nonumber \\
v_r \in [v_{r}^{{\rm min}},v_{r}^{{\rm max}}]
\end{eqnarray}  
the bounds may account for both thermodynamic considerations and functional aspects. A successuful computational framework in order to study the capabilities of the network  
is flux balance analysis (FBA): this consists on the maximization of linear objective function rappresenting e.g. the biomass production with the use of linear programming\cite{19}.
If a clear objective function is lacking 
and in order to retrieve general statistical properties of the network, 
a recent interesting issue regards the unbiased sampling of feasible flux states from the constraints (\ref{mb}). 
The network we consider is composed of $95$ reactions among $72$ compunds, upon considering the bounds given with the model and excluding leaves we are left with $86$ reactions among $68$ metabolites, the dimension of the underlying polytope is $D=23$. In order to analyze the performance of our method applied to this network, 
we have first to perform a preprocessing that eliminates ill-conditioning upon reducing the factor $R/r$: we have employed
the rounding technique exposed in \cite{articoloround}. 
In fig. 4 (left)we show the integrated autocorrelation times calculated by binning the data for the observables of interest, i.e. the reaction fluxes, 
for the normal algorithm and for our overelaxed version. The latter shows decreased autocorrelation times at all scales by a factor $15\%$  
We have then thus measured the time the point employs to escape from the angle nearby the origin along the same lines of the test performed in simpleces, the results are shown in fig. 4 (right) where it is possible to see
that the average escaping time for the overelaxed dynamics is $18\%$ lower.
\begin{figure}[h!]\label{fig3}
\begin{center}
\includegraphics*[width=.23\textwidth,angle=0]{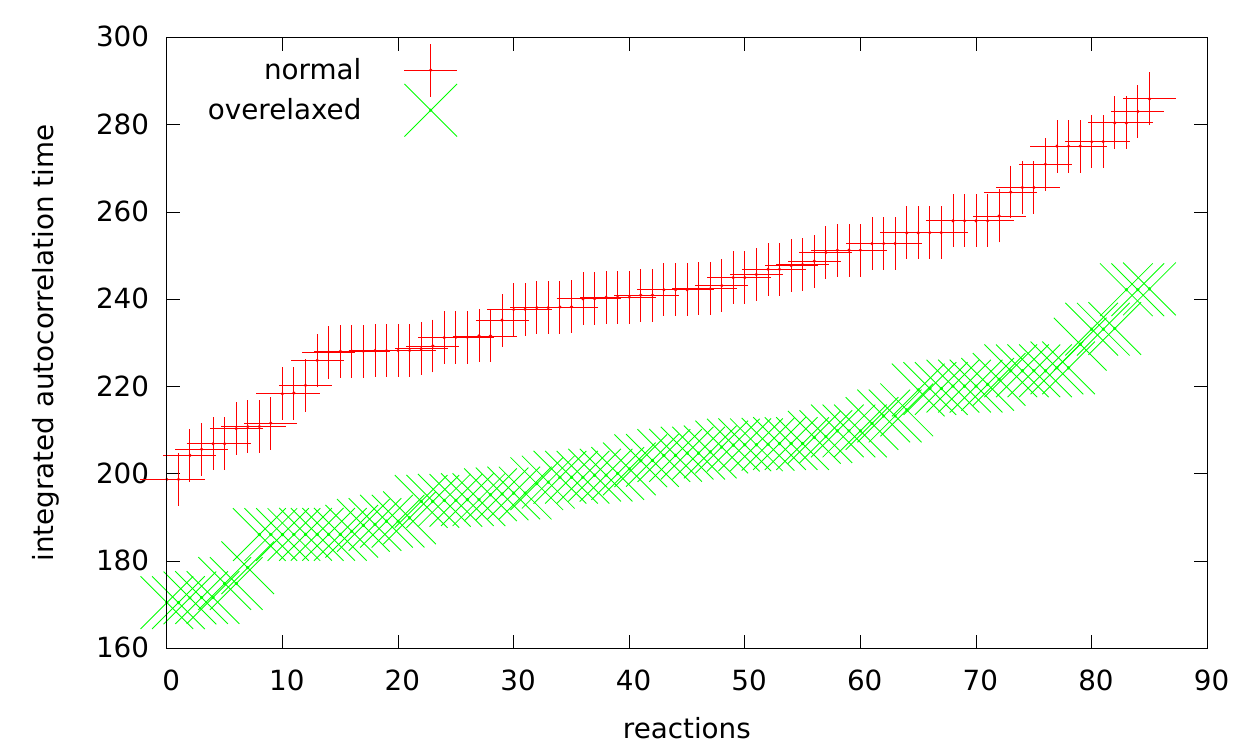}
\includegraphics*[width=.23\textwidth,angle=0]{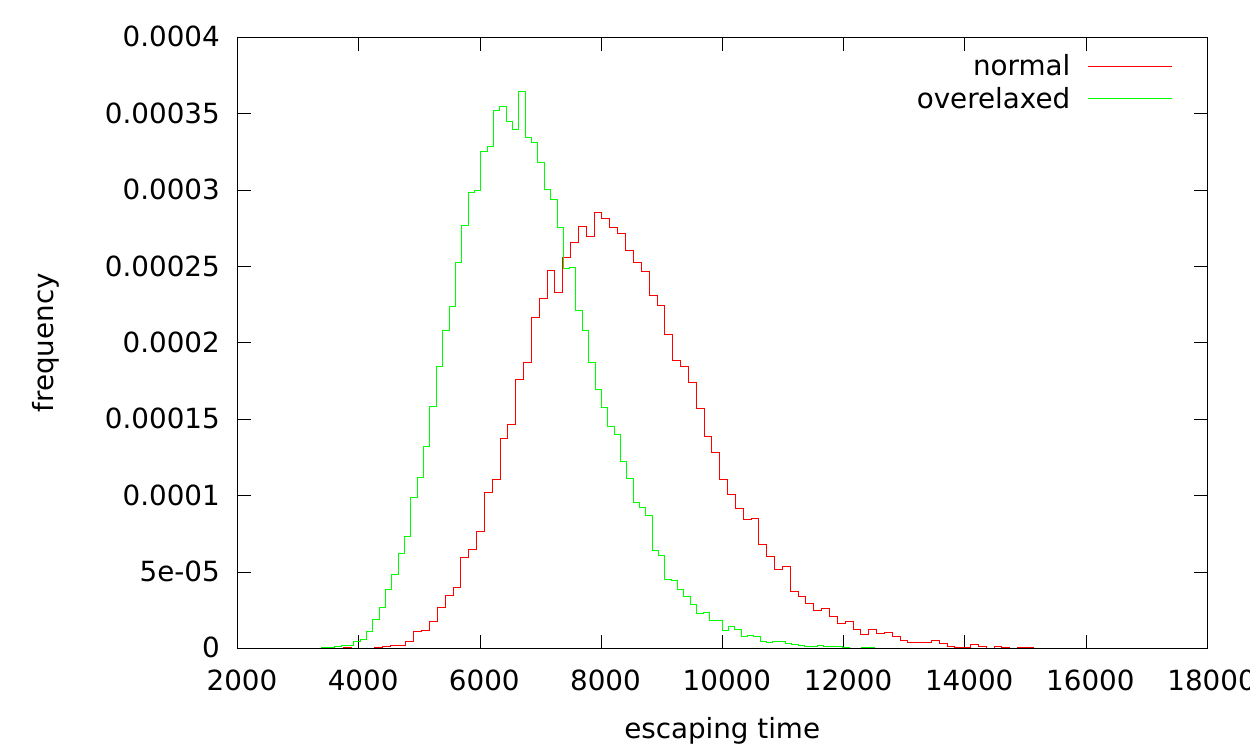}
\caption{Left: Integrated autocorrelation times for the hit-and-run dynamics normal and overelaxed for the reactions fluxes of the metabolic network model analyzed. 
The times for the overelaxed dynamics are on average $15\%$ lower. Right: Escaping time from a corner in the origin of  hit-and-run dynamics normal and overelaxed 
calculated for the the metabolic network model analyzed. The average time for the overelaxed dynamics is $18\%$ lower.}
\end{center}
\end{figure}

\section{Conclusions}
In this article a modification of the hit-and-run montecarlo algorithm has been proposed for the uniform sampling of convex regions in high dimensional spaces. 
Inspired by optimized montecarlo techniques used in statistical mechanics we propose to overelax the dynamics of the hit-and-run.
In this way it is possible to improve one of the  bottlenecks of the simple hit and run algorithm that is the diffusion from narrow corners. 
This permits to hasten the inference from uniform priors given by linear constraints and an example 
in the field of constraint-based modeling of cell metabolism was given.  The work can be extended in several interesting directions. 
First of all it would be worth to study more rigorously the markov chain defined by the method in order  to give bounds to the mixing time.
Then the method could be extended by considering non-linear conditional marginal probabilities. 
Finally it is worth noticing that an interesting  parallell approach to the problem of the uniform sampling 
with the use of marginalization algorithms like cavity methods has been recently proposed in particular in the field of metabolic networks flux sampling\cite{22,23}.
However, they work under the approximation of a tree-like network and are not guaranteed in general to converge to an uniform distribution, an issue that could be tested with
hit-and-run techniques like the one proposed in this article.

\appendix*
\section{Derivation of formula}
Consider a conditional probability over the segment that is linear in $t$
\begin{equation}
P(t|t_0) = f(t_0) t + g(t_0).
\end{equation}
Normalization implies that $f(t_0) \frac{L^2}{2}+g(t_0)L =1$,  from which 
\begin{equation}
g(t_0) = \frac{1}{L} - f(t_0)\frac{L}{2}.
e\end{equation}
In order to satisfy detailed balance we impose the simmetry $P(t|t_0)= P(t_0|t)$, and then we have $f(t_0) t + g(t_0) = f(t) t_0 + g(t)$, from which we have finally 
\begin{equation}
f(t_0) =\beta (t_0-L/2).
\end{equation}
It is possible to calculate the dependence of the correlation upon $\beta$:
we have $\langle t t_0 \rangle = \frac{1}{L}\int  dt_0 \langle t|t_0\rangle t_0$, where $\langle t |t_0\rangle = L/2+\beta(t_0-L/2)L^3/12$ , 
finally we have 
\begin{equation}
\frac{\langle t t_0 \rangle-\langle t \rangle^2}{\langle t^2 \rangle - \langle t \rangle^2}=\beta L^3/12.
\end{equation}
Since  $P(t|t_0)\geq 0$  $\forall t,t_0\in [0,L]$, we have 
\begin{equation}
\beta \geq
  \begin{cases}
    -\frac{1}{L(t-L/2)(t_0-L/2)} & \text{if } (t-L/2)(t_0-L/2)>0 \\
    \frac{1}{L(t-L/2)(t_0-L/2)}     & \text{if } (t-L/2)(t_0-L/2)<0
  \end{cases}
\end{equation}
from  which the value that maximizes the  anticorrelation is $\beta =-4/L^3$.  
\begin{acknowledgments}
The authors warmly thank prof E.Marinari for his constant and valuable guide during the development of this work. 
This work is supported by the DREAM Seed Project of the Italian Institute of Technology (IIT). The IIT Platform Computation is gratefully acknowledged.
\end{acknowledgments}

\end{document}